\documentclass[a4paper]{PoS}

\usepackage{amsmath}
\usepackage{subfigure}

\title{Dark Matter and Gauge Coupling Unification in Non-SUSY SO(10)
Grand Unified Models} 

\ShortTitle{DM and Gauge Coupling Unification in Non-SUSY SO(10) GUTs}

\author{\speaker{Natsumi Nagata}%
        \thanks{Based on work in collaboration with Yann Mambrini, Keith
        A. Olive, J\'{e}r\'{e}mie Quevillon, and Jiaming Zheng
        \cite{Mambrini:2015vna, Nagata:2015dma}.}\\
	William I. Fine Theoretical Physics Institute, School of Physics
        and Astronomy, \\University of Minnesota, Minneapolis, Minnesota
        55455, USA\\ 
	Kavli IPMU (WPI), University of Tokyo, Kashiwa, Chiba 277-8584,
        Japan \\ 
       	E-mail: \email{nagat006@umn.edu}}

\abstract{The stability of dark matter is naturally explained if there
is an additional U(1) symmetry which is spontaneously broken to a
discrete symmetry at a high-energy scale. Such a framework is realized
in the context of the SO(10) grand unification. In this work, we discuss
dark matter models in the non-supersymmetric SO(10) grand unified models
in which the stability of dark matter is assured by this mechanism. We
find that the requirement of gauge coupling unification with a
sufficiently high unification scale to evade the proton decay
constraints plays an important role in selecting viable candidates. Some
of the dark matter models can be tested in future dark matter direct
searches and proton decay experiments. 
}

\FullConference{18th International Conference From the Planck Scale to
		the Electroweak Scale \\ 
		25-29 May 2015\\
		Ioannina, Greece }

\begin{document}

%%%%%%%%%%%%%%%%%%%%%%%%%%%%%%%%%
\section{Introduction}
%%%%%%%%%%%%%%%%%%%%%%%%%%%%%%%%%

A variety of cosmological observations has provided strong evidences for
the existence of dark matter (DM). One of the distinct properties of dark
matter is its stability; for a particle to be a good DM candidate, it
should be stable or have a lifetime longer than the age of the
Universe. To insure the stability, it is often assumed that there exists
a discrete symmetry which prevents a DM particle from decaying into the
Standard Model (SM) particles. For instance, the $R$-parity in the
minimal supersymmetric (SUSY) SM allows the lightest SUSY particle
to be a good DM candidate. It is quite
often the case, however, that the origin of such a discrete symmetry is
unclear and thus its introduction seems more or less ad hoc. Thus, it is
of great interest if we can find some mechanism to generate a discrete
symmetry that makes DM stable.  

In fact, we can easily obtain a discrete symmetry if there is an
additional U(1) gauge symmetry beyond the SM gauge symmetries which is
spontaneously broken at a high-energy scale. Suppose that there is a
U(1) gauge symmetry for which generic fields (including SM fields)
$\varphi_i$ and a Higgs field $\varphi_H$ have charges $Q_i$ and $Q_H$,
respectively. Here, we normalize these charges so that they are
integers. We further assume that the charge of the Higgs field
$\varphi_H$ satisfies $Q_H \equiv 0$ (mod.~$N$) with $N$ being a
positive integer. A vacuum expectation value (VEV) of this Higgs field
$\langle \varphi_H \rangle$ breaks the extra U(1) symmetry. An important
observation here is that both the Lagrangian of the theory and the VEV
$\langle \varphi_H \rangle$ are invariant under the following
transformations: 
\begin{equation}
 \varphi_i \to \exp \left(\frac{i 2\pi Q_i}{N}\right) \varphi_i~, ~~~~~~
 \langle \varphi_H \rangle \to
 \exp \left(\frac{i 2\pi Q_H}{N}\right) 
\langle \varphi_H \rangle
= \langle \varphi_H \rangle ~.
\end{equation}
This indicates that $\langle\varphi_H\rangle$ breaks the extra U(1)
symmetry into a $\mathbb{Z}_N$ symmetry \cite{Kibble:1982ae,
Krauss:1988zc, Ibanez:1991pr, Ibanez:1991hv, Martin:1992mq}.

We can consider such an extra U(1) symmetry within the context of the
SO(10) grand unification \cite{Georgi:1974my, Fritzsch:1974nn}, as
SO(10) is a rank-five gauge group. Indeed, the SO(10) grand unified
theories (GUTs) have various attractive features. The SM quarks
and leptons as well as three right-handed neutrinos are embedded into
three generations of {\bf 16} spinor representations. Since SO(10) gauge
theory is free from gauge anomalies, the anomaly cancellation in the SM
can naturally be explained. In addition, SO(10) GUTs
can realize gauge coupling unification even without SUSY with the aid of
intermediate gauge symmetries (for a review of non-SUSY SO(10) GUTs with
intermediate gauge symmetries, see Ref.~\cite{Fukugita:1993fr}). In this
case, masses of right-handed neutrinos are of the order of the
intermediate gauge symmetry breaking scale $M_{{\rm int}}$, which
explains small neutrino masses via the seesaw mechanism
\cite{Minkowski:1977sc, Yanagida:1979as, GellMann:1980vs, Glashow:1979nm,
Mohapatra:1979ia,  Schechter:1980gr} if $M_{{\rm int}}$
is sufficiently high. 

In the non-SUSY SO(10) GUTs, the extra U(1) symmetry is broken at the
intermediate scale $M_{\rm int}$. Thus, by appropriately choosing the
Higgs field that breaks the intermediate gauge symmetry, we can obtain a
remnant discrete symmetry at low energies according to the mechanism
discussed above. In fact, it turns out that if a VEV of a ${\bf 126}$
({\bf 672}) of SO(10) breaks the intermediate gauge symmetry, then a
$\mathbb{Z}_2$ $(\mathbb{Z}_3)$ symmetry appears below the intermediate
scale \cite{DeMontigny:1993gy}. In particular, if we restrict ourselves
to consider SO(10) multiplets whose dimensions are $\leq 210$, then the
$\mathbb{Z}_2$ symmetry is the only possibility to be realized in such a
framework. This $\mathbb{Z}_2$ symmetry is found to be equivalent to the
so-called matter parity, $P_M =(-1)^{3(B-L)}$; this is not surprising
as SO(10) contains the $B-L$ symmetry. Since the SM fermions have the
matter parity odd and the SM Higgs field has the matter parity even, a
boson (fermion) with the matter parity odd (even) cannot decay into the
SM particles in the presence of this $\mathbb{Z}_2$ symmetry. Thus,
SO(10) GUTs can nicely explain the stability of DM with this remnant
$\mathbb{Z}_2$ symmetry \cite{Kadastik:2009dj, Kadastik:2009cu,
Frigerio:2009wf, Hambye:2010zb}.

In this work, we systematically study possibilities of DM in SO(10) GUT
models which is stabilized with this mechanism. We find
that two classes of DM candidates can be realized in this setup; one is
the non-equilibrium thermal DM (NETDM) \cite{Mambrini:2013iaa} and the
other is the weakly-interacting massive particles (WIMPs). The former DM
is a SM singlet fermion which is non-thermally produced from the
scattering of SM particles in the thermal bath through the
intermediate/GUT scale particle exchange. We discuss this case in the
next section following the study in Ref.~\cite{Mambrini:2015vna}. For
the latter case, DM can be either a scalar or a fermion, and has a
sizable interaction with the SM particles so that it can be thermalized
in the early Universe. The WIMP DM in SO(10) GUTs is thoroughly
discussed in Ref.~\cite{Nagata:2015dma}, and we briefly review the
result in Sec.~\ref{sec:wimpdm}. We focus on SO(10) GUT models in which
gauge coupling unification is achieved with a sufficiently high GUT
scale to evade constraints coming from proton decay experiments. We also
check whether these models yield the correct DM density and account for
small neutrino masses.

Throughout this article, we consider the following symmetry-breaking
chain,
\begin{equation}
 {\rm SO}(10) \longrightarrow G_{\rm int}
\longrightarrow {\rm SU}(3)_C \otimes {\rm SU}(2)_L \otimes {\rm U}(1)_Y
\otimes \mathbb{Z}_2 ~,
\label{eq:breakingchain}
\end{equation}
where the SO(10) GUT group is broken at the GUT scale $M_{\rm GUT}$ into
the intermediate gauge group $G_{\rm int}$ by a VEV of the GUT-scale
Higgs multiplet $R_1$. This intermediate gauge group is subsequently broken
to the SM gauge group at the intermediate scale $M_{\rm int}$ by a VEV
of the intermediate-scale Higgs multiplet $R_2$. The GUT and
intermediate scales are determined using the condition of gauge coupling
unification. We denote the DM multiplet by $R_{\rm DM}$. Among the
components in $R_{\rm DM}$, only the DM field has a mass much lighter
than the intermediate scale so that it can explain the observed DM
density, while the other components are supposed to have masses of
${\cal O}(M_{\rm int})$. Such a mass spectrum is obtained by fine-tuning
of the coefficients in the original Lagrangian terms, just like the
ordinary doublet-triplet splitting in the Higgs sector. We only consider
SO(10) multiplets whose dimensions are equal to or less than 210 in the
following discussion.

%%%%%%%%%%%%%%%%%%%%%%%%%%%%%%%%%%%%
\section{Singlet Fermion DM: NETDM}
%%%%%%%%%%%%%%%%%%%%%%%%%%%%%%%%%%%%

First, we discuss the case of SM singlet fermion DM. This class of DM
candidates cannot couple to the SM sector at renormalizable
level. Therefore, if new physics appears at a very high scale ($M_{\rm
int}$ in the present setup), the DM interactions with SM particles are
so weak that this DM cannot be in thermal equilibrium in the early
Universe. Nevertheless, such a DM particle can be produced via the NETDM
mechanism \cite{Mambrini:2013iaa} as we discuss below.

%%%%%%%%%%%%%%%%%%%%% TABLE %%%%%%%%%%%%%%%%%%%%%%%%%%%%%%%%%%%%%%%%%%
\begin{table}[t]
 \begin{center}
%\vspace{5pt}
\begin{tabular}{l|cc}
\hline
\hline
 & Model I& Model II \\
\hline
$G_{\text{int}}$ &$\text{SU}(4)_C\otimes \text{SU}(2)_L\otimes
     \text{SU}(2)_R$ &$\text{SU}(4)_C\otimes \text{SU}(2)_L\otimes
	 \text{SU}(2)_R\otimes D$\\ 
$R_{\text{DM}}$ & $({\bf 1}, {\bf 1}, {\bf 3})_D$ in ${\bf 45}_D$
& $({\bf 15}, {\bf 1}, {\bf
	 1})_W$ in ${\bf 45}_W$ \\
$R_1$ & ${\bf 210}_R$ & ${\bf 54}_R$ \\
$R_2$ & $({\bf 10}, {\bf 1}, {\bf 3})_C\oplus ({\bf 1}, {\bf 1}, {\bf
     3})_R$ & $({\bf 10}, {\bf 1}, {\bf 3})_C\oplus ({\bf 10}, {\bf 3},
	 {\bf 1})_C\oplus ({\bf 15}, {\bf 1}, {\bf
     1})_R$ \\
$M_{\text{int}}$ & $1.2\times 10^8$~GeV & $4.6\times 10^{13}$~GeV \\
$M_{\text{GUT}}$ & $4.4\times 10^{15}$~GeV & $7.4\times 10^{15}$~GeV \\ 
$g_{\text{GUT}}$ & $0.53$ & $0.57$ \\
\hline
\hline
\end{tabular}
\caption{Singlet fermion DM models \cite{Mambrini:2015vna}.}
\label{tab:model1and2}
 \end{center}
\end{table}
%%%%%%%%%%%%%%%%%%%%%%%%%%%%%%%%%%%%%%%%%%%%%%%%%%%%%%%%%%%%%%%%%%%%%%%

%%%%%%%%%%%%%% FIGURE %%%%%%%%%%%%%%%%%%%%%%%%%%%%%%%%%%%%
\begin{figure}[t]
\begin{center}
\subfigure[Model I]
 {\includegraphics[clip, width = 0.48 \textwidth]{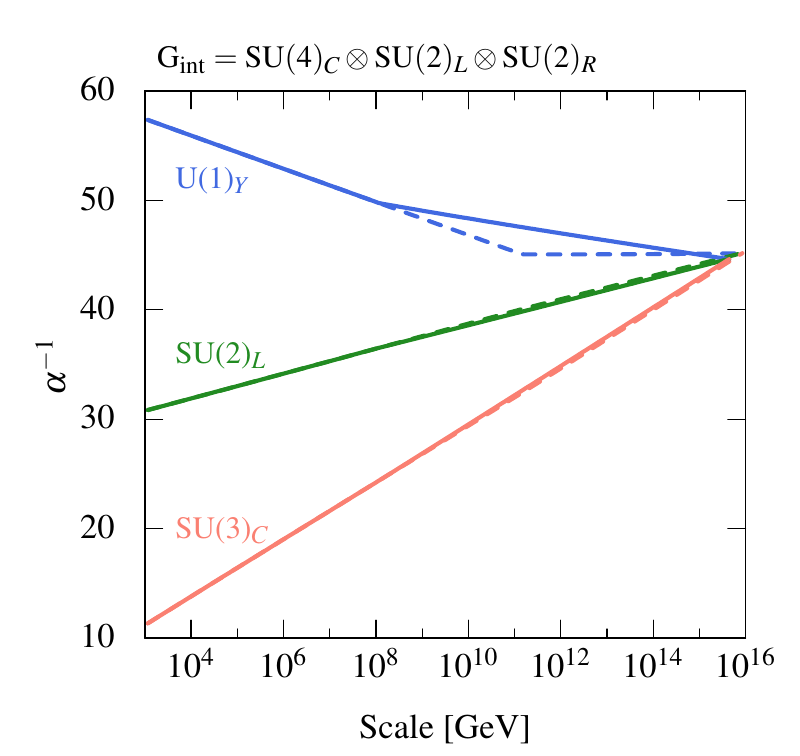}
 \label{fig:mod1}}
%\hspace{0.1\textwidth}
\subfigure[Model II]
 {\includegraphics[clip, width = 0.48 \textwidth]{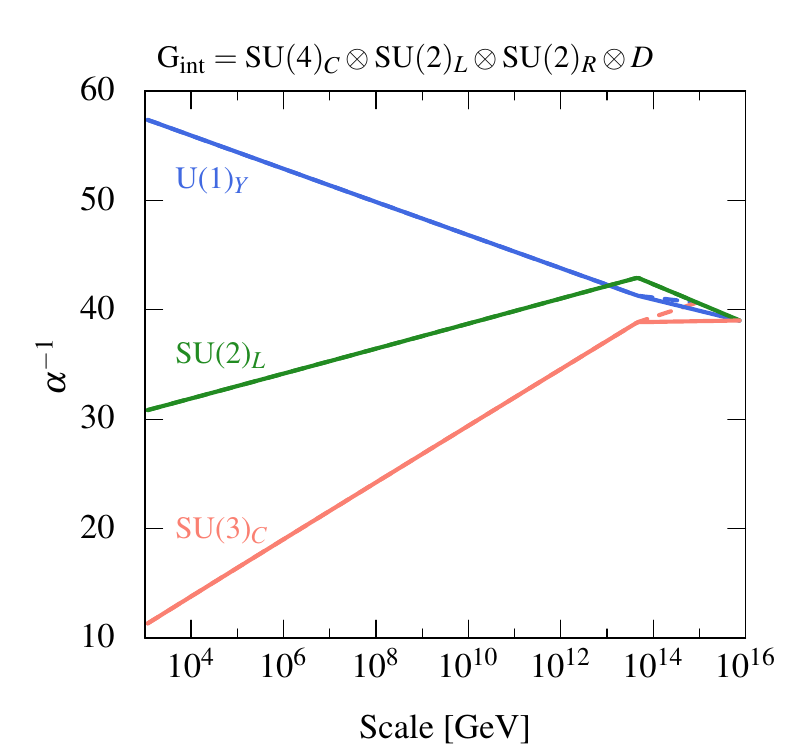}
 \label{fig:mod2}}
\caption{Running of gauge couplings \cite{Mambrini:2015vna}. Solid
 (dashed) lines show the case with (without) DM and additional Higgs
 bosons. Blue, green, and red lines represent the running of the
 U(1)$_Y$, SU(2)$_L$ and SU(3)$_C$ gauge couplings, respectively.}
\label{fig:gaugerun}
\end{center}
\end{figure}
%%%%%%%%%%%%%%%%%%%%%%%%%%%%%%%%%%%%%%%%%%%%%%%%%%%%%%%%%%

For a fermion to be stabilized by the $\mathbb{Z}_2$ symmetry in
Eq.~\eqref{eq:breakingchain}, it should have an even parity under the
symmetry. It is found that a SM singlet fermion which has an even
$\mathbb{Z}_2$ charge appears in a ${\bf 45}$, ${\bf 54}$, ${\bf 126}$,
or ${\bf 210}$ of SO(10). By requiring gauge coupling unification with a
sufficiently high GUT scale, we have found two promising models which
contain singlet fermion DM.\footnote{Here, we have assumed the DM
multiplet to be charged under the intermediate gauge symmetry. In this
case, the DM can be produced via the exchange of the intermediate-scale
particles. We can also consider the case where the DM multiplet is also
singlet under the intermediate gauge symmetry. This DM can still be
produced via the GUT-scale interactions. Such a possibility is also
discussed in Ref.~\cite{Mambrini:2015vna}.} We summarize the contents of
these models in 
Table~\ref{tab:model1and2}. One of the models has an intermediate gauge
symmetry of $G_{\text{int}} = \text{SU}(4)_C\otimes \text{SU}(2)_L\otimes
\text{SU}(2)_R$, while the other has $G_{\text{int}} =
\text{SU}(4)_C\otimes \text{SU}(2)_L\otimes \text{SU}(2)_R \otimes D$
with $D$ representing the left-right parity---an additional discrete
symmetry with respect to the interchange of left- and right-handed
fields. We refer to the former (latter) as Model I (II) in what
follows. In Model I, the singlet fermion DM stays inside the $({\bf 1},
{\bf 1}, {\bf 3})_D$ component of a {\bf 45}$_D$ of SO(10), while in Model
II it is in the $({\bf 15}, {\bf 1}, {\bf 1})_W$ of a ${\bf 45}_W$. Here,
the subscripts $R$, $C$, $W$, and $D$ indicate a real scalar boson, a
complex scalar boson, a Weyl fermion, and a Dirac fermion,
respectively. The $({\bf 10}, {\bf 1}, {\bf 3})_C$ component appearing
in $R_2$ originates from a ${\bf 126}_C$, whose VEV yields the remnant
$\mathbb{Z}_2$ symmetry. The other Higgs fields in $R_2$ are introduced
to give masses of ${\cal O}(M_{\text{int}})$ to all of the components in
$R_{\text{DM}}$ except the DM field. The GUT and intermediate scales as
well as the unified gauge coupling $g_{\text{GUT}}$ in
Table~\ref{tab:model1and2} are evaluated by using the two-loop
renormalization group equations (RGEs). We also show the running of the
gauge couplings in these models in Fig.~\ref{fig:gaugerun}. Here, the
solid and dashed lines show the cases with and without DM and additional
intermediate-scale Higgs multiplets, respectively. The U(1)$_Y$ gauge
coupling constant above the intermediate scale is defined by
$\frac{1}{\alpha_Y} \equiv \frac{3}{5} \frac{1}{\alpha_{2R}}
+\frac{2}{5}\frac{1}{\alpha_4}$ where $\alpha_{2R}$ ($\alpha_4$) is the
SU(2)$_R$ (SU(4)$_C$) gauge coupling, while the SU(3)$_C$ gauge coupling
above $M_{\rm int}$ is given by $\alpha_3 \equiv \alpha_4$. As can be
seen from the plots, the presence of the DM and extra intermediate-scale
Higgs multiplets significantly affects the running of the gauge coupling
constants.

%%%%%%%FIGURE%%%%%%%%%%%%%%%%%%%%%%
\begin{figure}[t]
\begin{center}
\includegraphics[height=75mm]{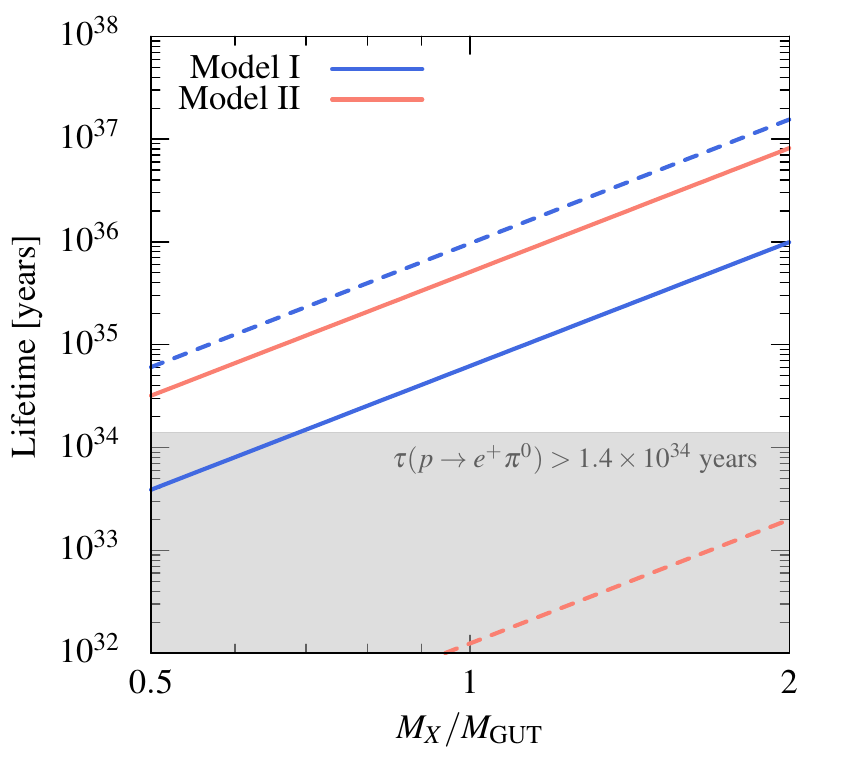}
\caption{{Proton lifetimes as functions of $M_X/M_{\text{GUT}}$
 \cite{Mambrini:2015vna}. Blue and red solid  lines correspond to 
 Model I and II, respectively.  Blue and red dashed lines
 represent the cases for  $G_{\text{int}} = \text{SU}(4)_C\otimes
 \text{SU}(2)_L \otimes \text{SU}(2)_R$ and $G_{\text{int}} =
 \text{SU}(4)_C\otimes \text{SU}(2)_L \otimes \text{SU}(2)_R \otimes D$,
 respectively, where the DM and extra Higgs multiplets are not included. 
 Shaded region shows the current experimental
 bound, $\tau (p\to e^+\pi^0)> 1.4 \times 10^{34}~{\rm years}$
 \cite{Babu:2013jba}. }}
\label{fig:protdec}
\end{center}
\end{figure}
%%%%%%%%%%%%%%%%%%%%%%%%%%%%%%%%%

As we have obtained the GUT scales in the models, we can now evaluate
proton decay lifetimes for each model. In the non-SUSY GUTs, proton
decay is induced by the exchange of the GUT-scale gauge bosons. The
dominant decay mode in this case is the $p\to e^+ \pi^0$ process. A
calculation of proton decay rates in SO(10) GUTs is presented in
Appendix~B in Ref.~\cite{Nagata:2015dma}, and we follow the prescription
given there. We show the resultant proton decay lifetimes in the models
in Fig.~\ref{fig:protdec}. Here, the 
blue solid and red solid lines represent proton lifetimes in Model I and
II, respectively, while the dashed lines correspond to the cases
in which the DM and extra Higgs multiplets are not included. Proton
lifetimes depend on the GUT-scale gauge boson mass $M_X$. Although $M_X$
is expected to be of the order of the GUT scale, its precise value
cannot be determined from low energies, and this ignorance therefore causes
uncertainty in the proton decay calculation. Considering this, we
estimate the uncertainty by varying $M_X$ from $M_{\text{GUT}}/2$ to
$2M_{\text{GUT}}$ in this plot. The gray shaded region in
Fig.~\ref{fig:protdec} shows the current experimental bound, $\tau (p\to
e^+\pi^0)> 1.4 \times 10^{34}~{\rm years}$ \cite{Babu:2013jba}. From
this figure, we find that proton lifetimes for $M_X \simeq
M_{\text{GUT}}$ lie above the current constraint in both of the
models. Notice that the presence of the DM and extra Higgs multiplets
gives a significant impact on the prediction of proton lifetimes;
indeed, in the absence of these particles, the $G_{\text{int}} =
\text{SU}(4)_C\otimes \text{SU}(2)_L \otimes \text{SU}(2)_R \otimes D$
case predicts too rapid proton decay rate to evade the current
experimental limit, but the addition of these particles makes Model
II viable. Anyway, in both of these models, proton decay lifetimes are
predicted to be rather short. Hence, these models may be probed in
future proton decay experiments.

Next, we consider neutrino masses in these models. In both of the
models, neutrino masses are given by the ordinary seesaw mechanism 
\cite{Minkowski:1977sc, Yanagida:1979as, GellMann:1980vs, Glashow:1979nm,
Mohapatra:1979ia, Schechter:1980gr}. An important caveat here is that in
the case of SO(10) GUTs Dirac Yukawa couplings for neutrinos are related to
other Yukawa couplings. In particular, since the SU(4)$_C$ symmetry is
manifest above $M_{\rm int}$ in our models and a VEV of the $({\bf 1},
{\bf 2}, {\bf 2})$ component, which is identified as the SM Higgs field
in the minimal setup, does not break the SU(4)$_C$ symmetry, the Dirac
masses of neutrinos are equal to up-quark masses $m_{u_i}$ up to
logarithmic corrections in the low-energy region. As a consequence,
neutrino masses $m_{\nu_i}$ are given by 
\begin{equation}
 m_{\nu_i} \simeq \frac{m_{u_i}^2}{M_{\text{int}}}~,
\label{eq:numassseesaw}
\end{equation}
where we have used the fact that Majorana masses for right-handed
neutrinos are ${\cal O}(M_{\text{int}})$. From the results given in
Table~\ref{tab:model1and2}, we then find that the Model II predicts
neutrino masses compatible with the present data, while those in Model I
are too large. This is because of a low intermediate scale in Model I,
as can be seen from Fig.~\ref{fig:mod1}. In this sense, Model II is more
favorable than Model I. A simple way to make Model I viable is to
exploit the $({\bf 15}, {\bf 2}, {\bf 2})$ component of the {\bf 126}
Higgs multiplet. This component contains an SU(2)$_L$ doublet Higgs
field, which can have a VEV of the order of the electroweak scale. Since
this VEV breaks the SU(4)$_C$ symmetry, one can choose its VEV and
Yukawa couplings such that its contribution to Dirac mass terms of
neutrinos cancels the above contribution. In this case, we can evade the
relation \eqref{eq:numassseesaw}, and desirable neutrino masses can be
obtained even with a low intermediate scale. We also find that the
presence of a $({\bf 15}, {\bf 2}, {\bf 2})$ field scarcely affects gauge
coupling unification, as it changes the beta functions of all of the
gauge coupling constants by similar amounts \cite{Bajc:2005zf}. Therefore, the
introduction of a $({\bf 15}, {\bf 2}, {\bf 2})$ resolves the neutrino
mass problem in Model I without largely modifying the GUT and
intermediate scales obtained above. 

%%%%%%%FIGURE%%%%%%%%%%%%%%%%%%%%%%
\begin{figure}[t]
\begin{center}
\includegraphics[height=50mm]{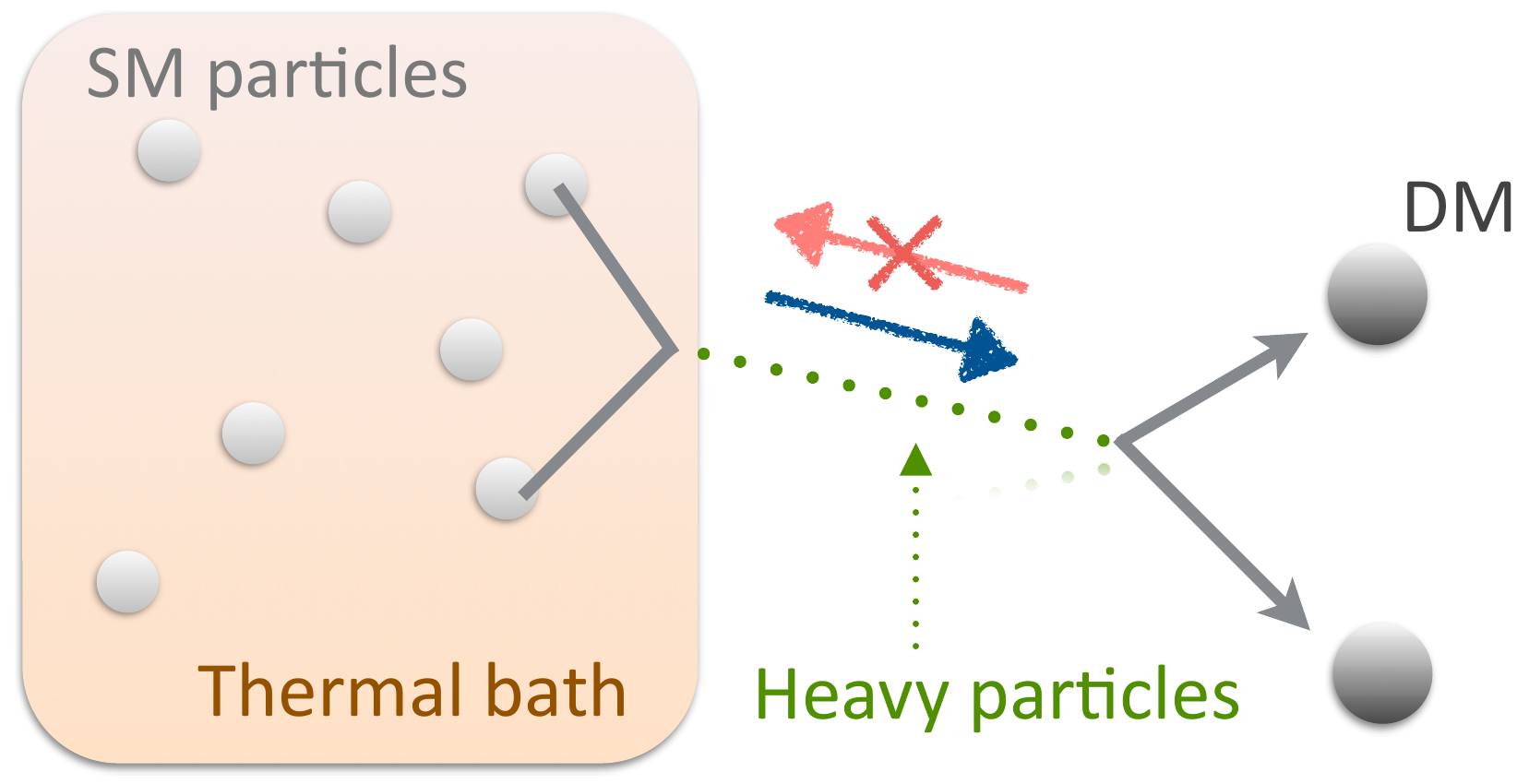}
\caption{{The NETDM mechanism \cite{Mambrini:2013iaa} for DM
 production. }}
\label{fig:NETDMfig}
\end{center}
\end{figure}
%%%%%%%%%%%%%%%%%%%%%%%%%%%%%%%%%

Now we consider the DM production mechanism in these models. As
mentioned above, DM particles in these models do not come into thermal
equilibrium in the early Universe as long as the reheating temperature
$T_{\rm RH}$ is lower than the intermediate scale. In this case, DM is
produced non-thermally from the scattering of the SM particles in the
thermal bath through the exchange of intermediate/GUT scale
particles. The pair annihilations of DM particles during the DM
production are negligible. This situation is illustrated in 
Fig.~\ref{fig:NETDMfig}. The current DM abundance in this case is
evaluated by using the following Boltzmann equation:
\begin{equation}
 \frac{d Y_{\text{DM}}}{dx} = \sqrt{\frac{\pi g_*}{45}}
M_{\text{DM}} M_{\text{Pl}} \frac{\langle \sigma v \rangle}{x^2} 
Y_{\text{eq}}^2 ~,
\label{eq:boltz}
\end{equation}
where $Y_{\rm DM} \equiv n_{\rm DM}/s$, $Y_{\rm eq} \equiv n_{\rm
eq}/s$, $n_{\text{DM}}$ is the number density of DM, $n_{\rm eq}$ is the
equilibrium number density of each individual initial state SM particle,
$s$ is the entropy of the Universe, $x\equiv M_{\text{DM}}/T$, $T$ is
the temperature of the Universe, $M_{\text{DM}}$ is the DM mass,
$M_{\text{Pl}} \equiv 1/\sqrt{G_N} = 1.22\times 10^{19}$~GeV, $g_*$ is
the effective degrees of freedom in the thermal bath, and $\langle
\sigma v \rangle$ denotes the thermally averaged total annihilation
cross section of the initial SM particles into the DM pair. Notice that
there is no self-annihilation contribution in the right-hand side in
Eq.~\eqref{eq:boltz}, as mentioned above. By solving this equation, we
can obtain DM relic abundance as a function of $M_{\text{DM}}$ and
$T_{\text{RH}}$.

%%%%%%%%%%%%%% FIGURE %%%%%%%%%%%%%%%%%%%%%%%%%%%%%%%%%%%%
\begin{figure}[t]
\begin{center}
\subfigure[Model I]
 {\includegraphics[clip, width = 0.48 \textwidth]{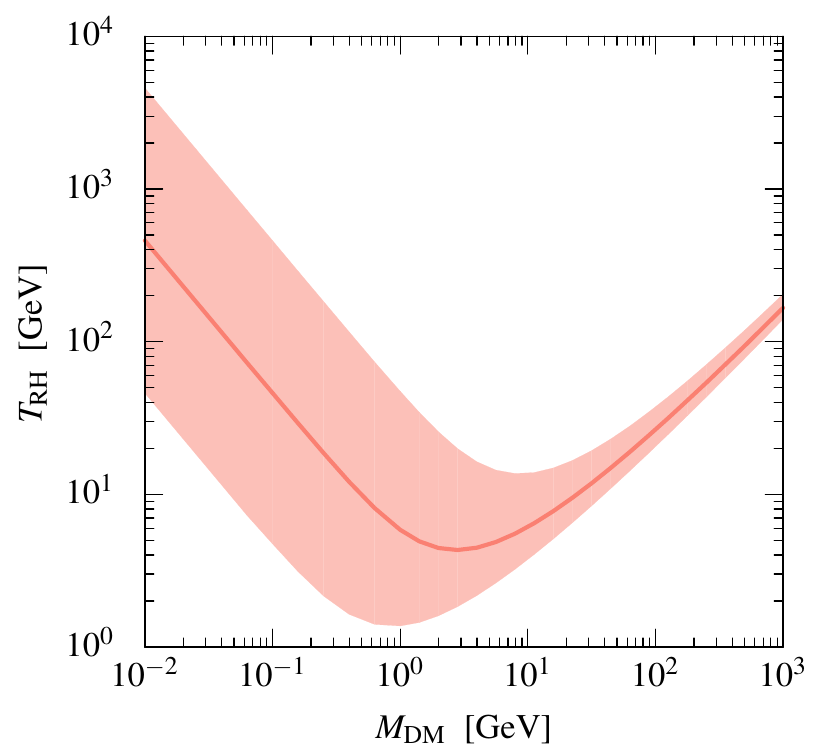}
 \label{fig:trhmod1}}
%\hspace{0.1\textwidth}
\subfigure[Model II]
 {\includegraphics[clip, width = 0.48 \textwidth]{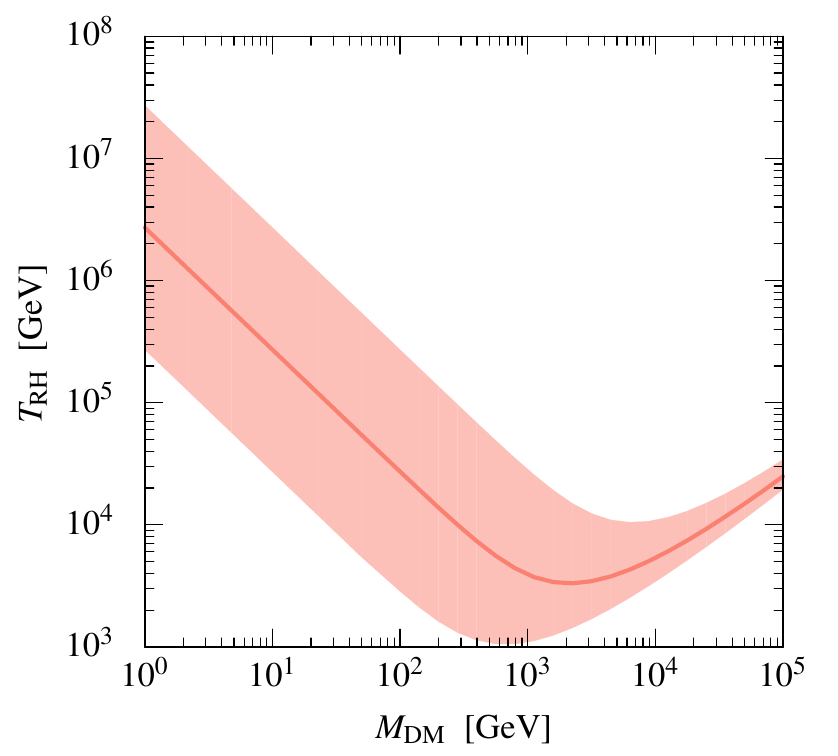}
 \label{fig:trhmod2}}
\caption{\it Reheating temperature as a function of
 $M_{\text{DM}}$ \cite{Mambrini:2015vna}. Pink band shows the
 theoretical uncertainty.}  
\label{fig:trh}
\end{center}
\end{figure}
%%%%%%%%%%%%%%%%%%%%%%%%%%%%%%%%%%%%%%%%%%%%%%%%%%%%%%%%%%

In Fig.~\ref{fig:trh}, we show the regions favored by the DM relic
abundance in the $M_{\rm DM}$--$T_{\rm RH}$ plane for each model. Here,
the pink bands indicate the uncertainty of our computation. We see that
Model I predicts low value of $T_{\rm RH}$ except for the fine-tuning
region $T_{\rm RH} \simeq M_{\text{DM}}$. For such a low reheating
temperature, baryogenesis is usually quite challenging. On the other
hand, $T_{\text{RH}}$ can be sufficiently high in the case of Model
II. Thus, Model II is again favored compared to Model I.

%%%%%%%%%%%%%%%%%%%%%%%%%%%%%%%%
\section{WIMP DM Candidates}
\label{sec:wimpdm}
%%%%%%%%%%%%%%%%%%%%%%%%%%%%%%%

Next, we briefly discuss WIMP DM in the SO(10) GUT models. For
systematic classification of such possibilities and detailed discussions
on the models, see Ref.~\cite{Nagata:2015dma}. In the case of WIMP DM,
we have both scalar and fermion candidates. For a scalar particle to be
stable, it should have odd matter parity, which can originate from
either a ${\bf 16}$ or ${\bf 144}$ of SO(10). By requiring a
sufficiently high unification scale to ensure a proton decay lifetime
compatible with the current experimental limits, we find that only the
SM singlet DM or the SU(2)$_L$ doublet with hypercharge $Y=\pm 1/2$
scalar DM
can be realized in the low-energy regions, with possible
intermediate scales being $\text{SU}(4)_C\otimes \text{SU}(2)_L \otimes
\text{SU}(2)_R$, $\text{SU}(3)_C\otimes \text{SU}(2)_L \otimes
\text{SU}(2)_R \otimes \text{U}(1)_{B-L}$, or $\text{SU}(3)_C\otimes
\text{SU}(2)_L \otimes \text{SU}(2)_R \otimes \text{U}(1)_{B-L} \otimes
D$. The former case, in which a singlet scalar is added to the SM as a
DM candidate, is one of the simplest extensions of the SM and has widely
been discussed so far \cite{Silveira:1985rk, McDonald:1993ex,
Burgess:2000yq}. The latter DM candidate is often called the inert Higgs
doublet DM \cite{Deshpande:1977rw, Ma:2006km, Barbieri:2006dq,
LopezHonorez:2006gr}. As for fermionic DM candidates, we obtain an
SU(2)$_L$ triplet $Y=0$ DM candidate from a ${\bf 45}$ of SO(10), though
it requires additional Higgs multiplets around the intermediate scale to
realize good gauge coupling unification. We can also
find SU(2)$_L$ doublet DM candidates. Such a DM candidate in
general requires extra Majorana fermions beyond the DM multiplet;
otherwise the DM candidate becomes a Dirac fermion and such a
possibility has already been excluded by direct detection experiments
since Dirac fermion DM with non-zero hypercharge has a large
DM-nucleon scattering cross section. To evade the constraint, the
SU(2)$_L$ doublet DM has to mix with the extra Majorana fermions to
be split into two pseudo-Dirac fermions. Moreover, to also suppress the
inelastic scattering of the DM with nucleons, the mass difference
between these pseudo-Dirac fermions should be $\gtrsim 100$~keV
\cite{Nagata:2014wma}. This condition gives an upper bound on the masses
of the extra Majorana fermions \cite{Nagata:2014aoa}, and if we assume
these masses to be ${\cal O}(M_{\text{int}})$, this bound leads to an
upper bound on $M_{\text{int}}$. Taking into account this constraint, as
well as that from proton decay bounds, we found several models for
the SU(2)$_L$ doublet DM, which turn out to have either $G_{\rm int} =
\text{SU}(4)_C\otimes \text{SU}(2)_L \otimes \text{SU}(2)_R$ or
$\text{SU}(4)_C\otimes \text{SU}(2)_L \otimes \text{U}(1)_R$. These
non-SUSY SO(10) WIMP DM models can be tested in future DM direct
searches and proton decay experiments.

%%%%%%%%%%%%%%%%%%%%%%%%%%%%%%%%%%%%
\section{Conclusion and Discussions}
%%%%%%%%%%%%%%%%%%%%%%%%%%%%%%%%%%%%

In this work, we have discussed DM candidates in non-SUSY SO(10) GUT
models with an intermediate gauge symmetry. In these models, the
stability of DM is achieved thanks to a remnant $\mathbb{Z}_2$ symmetry
which is a subgroup of SO(10). The DM can be either NETDM or WIMP
DM. For both cases, the requirement of gauge coupling unification with a
sufficiently high GUT scale to evade the proton decay limit plays a
significant role in selecting viable candidates. In the case of NETDM,
we found two promising models which satisfy the above condition. The
DM particles in these models are non-thermally produced from the
scattering of the SM particles in the thermal bath via the
exchange of intermediate-scale particles; we have computed the favored
reheating temperature which yields the correct DM density. As for the
WIMP DM case, we have found models that include a SM singlet, an
inert Higgs doublet, an SU(2)$_L$ triplet, or an SU(2)$_L$ doublet
DM candidate. In any cases, future DM direct searches and proton decay
experiments can probe these DM models.

Finally, we would like to comment on some previous studies. In
Refs.~\cite{Kadastik:2009dj, Kadastik:2009cu}, a SM singlet scalar and
SU(2)$_L$ doublet scalar DM originating from a {\bf 16} of SO(10) are
considered, where SO(10) is assumed to be broken to $\text{SU}(5)\otimes
{\rm U}(1)_X$ and this U(1)$_X$ is broken to the $\mathbb{Z}_2$
symmetry. Gauge coupling unification and the proton decay constraints
are not discussed in detail. In Ref.~\cite{Frigerio:2009wf}, on the
other hand, both scalar and fermion DM candidates are
discussed. Intermediate gauge symmetries are not discussed there, and
for this reason, an SU(2)$_L$ triplet with $Y=0$ is found to be the only
promising candidate for DM  which may make the unification scale high
enough, though gauge coupling unification is challenging if one focuses on
minimal field contents. 

\acknowledgments
The work of the author was supported by Research Fellowships of the
Japan Society for the Promotion of Science for Young Scientists.

%%%%%%%%%%%%%%%%%%%%% References %%%%%%%%%%%%%%%%%%%%%%%%%%%%%%%
\bibliographystyle{JHEP}
\bibliography{ref}
%%%%%%%%%%%%%%%%%%%%%%%%%%%%%%%%%%%%%%%%%%%%%%%%%%%%%%%%%%%%%%%%

\end{document}